# Turning ZrTe$_5$ into semiconductor through atomic intercalation

Qi-Yuan Li[1], Yang-Yang Lv[2], Jinghui Wang[1], Song Bao[1], Wei Shi[1], Li Zhu[1], Wei-Min Zhao[1], Cheng-Long Xue[1], Zhen-Yu Jia[1], Libo Gao[1,3], Y. B. Chen[1,3], Jinsheng Wen[1,3], Yan-Feng Chen[2,3] and Shao-Chun Li[1,3]*

*1 National Laboratory of Solid State Microstructures, School of Physics, Nanjing University, Nanjing 210093, China*

*2 National Laboratory of Solid State Microstructures, Department of Materials Science and Engineering, Nanjing University, Nanjing 210093, China*

*3 Collaborative Innovation Center of Advanced Microstructures, Nanjing University, Nanjing 210093, China*

* Email: scli@nju.edu.cn

**Abstract**

In this work, we use the liquid ammonia method to successfully intercalate potassium atoms into ZrTe$_5$ single crystal, and find a transition from semimetal to semiconductor at low temperature in the intercalated ZrTe$_5$. The resistance anomalous peak is gradually suppressed and finally disappears with increasing potassium concentration. Whilst, the according sign reversal is always observed in the Hall resistance measurement. We tentatively attribute the semimetal-semiconductor transition to the lattice expansion induced by atomic intercalation and thereby a larger energy band gap.

**Introduction**

Study of Zirconium pentatelluride (ZrTe$_5$) was initiated about three decades ago owing to its potential in thermoelectricity and the resistance anomaly etc [1-8]. Recently, ZrTe$_5$ has attracted increasing attentions because of its mysterious topological nature. DFT calculation predicted that the single-layer ZrTe$_5$ is an ideal two-dimensional topological insulator (2D TI) hosting an indirect band gap of ~0.1 eV [9], but its bulk counterpart resides in proximity of the boundary of topological phase transition from weak to strong TI, through a 3D Dirac semimetal state [9-11]. Even though great efforts have been devoted to experimentally verify the topological characteristics, the topology of bulk ZrTe$_5$ is still under heavy debate. Either weak TI, strong TI or Dirac semimetal state, has ever been proposed based on the experimental observations with different techniques [12-32]. The bulk band gap and the topological edge state were detected by ARPES and by STM [14, 15, 29], which indicates that ZrTe$_5$ is a weak TI. Controversially, signatures of 3D Dirac semimetal were also demonstrated in magnetoresistance [18-20, 32] and magneto-infrared spectroscopy measurements [12, 13, 17, 24, 25]. The strong TI of ZrTe$_5$ has been also proposed [21, 27]. In addition, more exotic physical properties have been also found in this material, such as Zeeman splitting [20, 33, 34], Chiral Magnetic effect (CME) [16], 3D Quantum Hall effect (QHE) [35], Anomalous Hall effect (AHE) [36], Giant Planar Hall Effect (PHE) [37], and Discrete Scale Invariance (DSI) [38, 39].

Considering the extreme sensitivity of the band structure of ZrTe$_5$ to the lattice constant, the above mentioned experimental divergence may be likely due to the difference in sample growth. It is worthwhile noting that the Dirac node of ZrTe$_5$ is not protected by the crystal symmetry, and a band gap can be opened by a perturbation to turn it into topological insulator. High Pressure was applied to tune ZrTe$_5$ into superconductivity[40], and to induce topological phase transition[41]. The exfoliated ZrTe$_5$ nanosheet shows metallic behavior and tunable resistance maximum, which can be ascribed to the thickness-dependent band structure [42, 43]. Up to now, most of the explored ZrTe$_5$ bulk samples, grown by CVT method, behave as metallic state at low temperature and the resistance maximum is located at ~130-150 K, which indicates that bulk states of ZrTe$_5$ contribute to the electrical conductivity. A recent study showed that such metallic state may be induced by the intrinsic Te defects induced during CVT growth, and an alternate Flux growth method may help reduce such defects and realize semiconductor at low temperature [44].

In this study, we successfully intercalate potassium atoms into ZrTe$_5$ single crystal by liquid ammonia route, which makes it possible to gradually tune the electronic structure of ZrTe$_5$. Electrical transport measurement shows a smooth evolution from metallic to insulating behavior at low temperature upon increasing the potassium concentration, which is accompanied by the suppression of resistance maximum. Raman spectroscopy measurement indicates that the van der Waals gap of ZrTe$_5$ is enlarged by intercalation, resembling to a negative pressure applied along the *b*-axis. Therefore, it can be understood that such metal-semiconductor transition is induced by the increased band gap through lattice expansion. Furthermore, Hall resistance measurement always shows the sign reversal between 120 K and 160 K, no matter whether the resistance peak is suppressed or not, implying that the anomalous resistance peak may not be necessary to accompany with the carrier type reversal.

**Experimental methods**

Single crystal ZrTe$_5$ was grown by chemical vapor transport (CVT) method with iodine (I$_2$) as the transport agent, as described in the previous report [45]. The K-intercalated ZrTe$_5$ sample was obtained through the liquid ammonia route [46, 47]. At first, the ZrTe$_5$ single crystal and the potassium sheet were placed into the quartz tube. Secondly, the quartz tube was then sealed and vacuumed, and then placed into a cold trap consisted of the alcohol / water mixture. High-purity ammonia gas (99.999%) was introduced into the quartz tube and condensed as liquid ammonia to immerse the samples. The potassium pieces were dissolved into liquid ammonia and intercalated into ZrTe$_5$ single crystal. After the intercalation, the liquid ammonia was pumped away. Electrical resistance measurements were carried out in a physical property measurement system (PPMS) from Quantum Design. The Raman spectroscopy was measured using a WITec alpha 300R with a laser of 532 nm.

**Results and Discussion**

The schematic illustration of the ZrTe$_5$ single crystal with K atoms intercalated in the van der Waals gap is shown in the inset to Fig. 1. ZrTe$_5$ is a quasi-two-dimensional layered material which has an orthorhombic crystal structure with the space group of *Cmcm* ($D_{2h}^{17}$) [1, 48]. In ZrTe$_5$ single-layer sheet, each Zr atom is surrounded by three Te atoms forming a ZrTe$_3$ triangular prism along the *a*-axis. These ZrTe$_3$ triangular prisms are linked by the parallel zigzag Te chains in the c-axis, thus forming a quasi-two-dimensional *a-c* plane. The bulk ZrTe$_5$ is formed by these sheets stacked along the *b*-axis via van der Waals interactions.

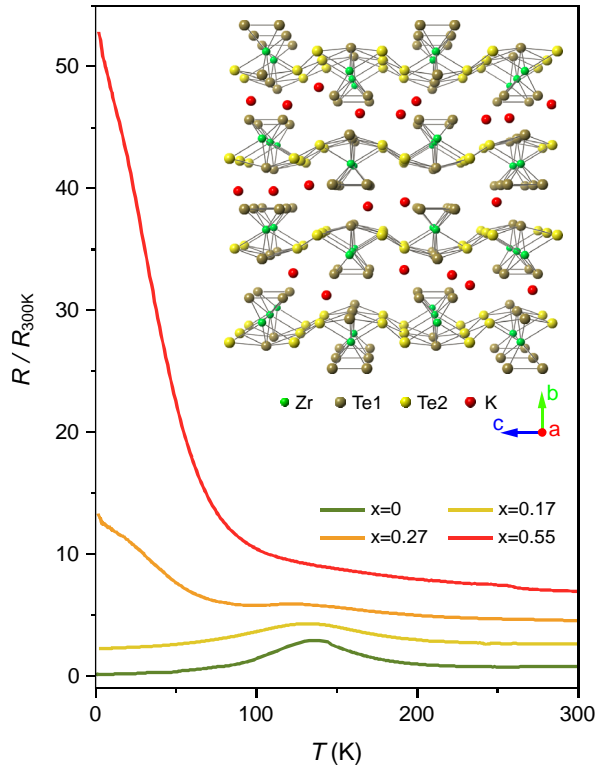

**Figure 1** (Color online) Resistance (R) as a function of temperature measured on the K intercalated K$_x$ZrTe$_5$. The R is normalized to the value at 300 K. The curves are shifted by 2 in the vertical direction for better clarity. The inset shows the schematic illustration of the ZrTe$_5$ single crystal intercalated with K atoms.

Figure 1 shows the temperature (*T*) dependence of normalized resistance (*R*)

measured over various $K_x ZrTe_5$ ($x$ = 0 to ~0.55) samples under zero magnetic field. The current was applied along the *a*-axis. For the pristine $ZrTe_5$ single crystal, the R-T curve exhibits an anomalous resistance peak around $T_p$ = ~135 K, and a typical metallic behavior at temperatures below $T_p$, in agreement with previous reports[45]. Surprisingly, the low temperature resistance is tremendously increased upon the K intercalation. As $x$ is greater than ~0.27, the sample undergoes a semimetal to semiconductor transition at low temperature. Meanwhile, the anomalous resistance peak is always persisted at ~135 K, but its intensity is gradually decreased and finally disappears when $x$ is greater than ~0.55.

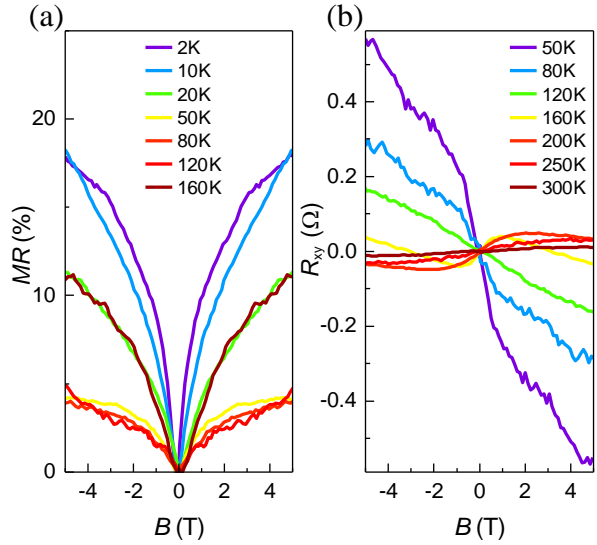

**Figure 2** (Color online) Magnetic field dependence of (a) the longitudinal magnetoresistance $MR = [R(H)/R(0)-1] \times 100\%$ and (b) the Hall resistance $R_{xy}$ measured on $K_{0.55}ZrTe_5$ at various temperatures.

To further investigate the influence of magnetic field to the transport properties, we performed the magneto-resistance measurement under the magnetic field of up to 5T applied along the *b*-axis. Figure 2(a) and 2(b) present the field dependence of longitudinal magnetoresistance $MR = [R(H)/R(0) - 1] \times 100\%$ and Hall resistance $R_{xy}$ measured on $K_{0.55}ZrTe_5$ at various temperatures. Due to the asymmetrical issue of electrodes during the measurement, the obtained results are symmetrized for both $MR$ and $R_{xy}$. Positive magnetoresistance effect is observed in the whole explored temperature range (from 2 K to 160 K) and magnetic field range (from 0 T to 5 T). It can be seen that *MR*(*H*) roughly follows a $R \sim \sqrt{B}$ dependence at low temperature, indicating the existence of weak antilocalization effect [49]. Furthermore, the temperature dependence of MR shares a similar trend as the intrinsic ZrTe$_5$, particularly the CVT sample [44]: The magnitude of *MR* drops first and then rises with increasing temperature, as shown in Fig 2(a). It is noteworthy that the resistance peak is completely suppressed as indicated in the *R-T* plot of Fig. 1. Even though the trend is similar, the positive MR effect in K$_x$ZrTe$_5$ is significantly weakened by about one and a half orders of magnitude. For instance, previous study showed the *MR* of ~900% at 5T for CVT ZrTe$_5$[44], while that of K$_{0.55}$ZrTe$_5$ is only ~17%. Therefore, it is suggested that the influence of the magnetic field on resistance is small but not negligible.

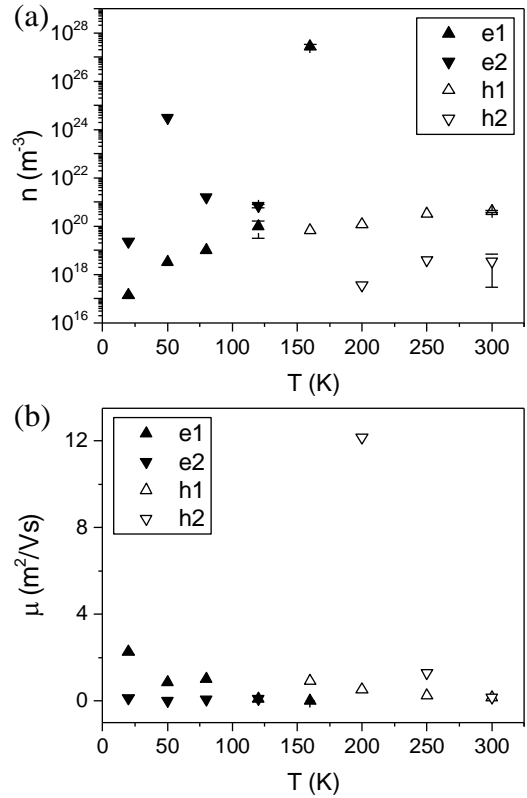

**Figure 3** Temperature dependence of carrier density (a) and mobility (b) for K$_{0.55}$ZrTe$_5$ extracted from the fitting to Hall conductivity.

As shown in Fig. 2(b), the Hall resistance $R_{xy}$ exhibits a negative slope at low temperatures of T ≤ 120-160 K, indicating the electron carriers dominated transport. As temperature is increased to ~160 K, the Hall resistance near zero field undergoes a sign reversal from negative to positive, consistent with the observation in the intrinsic ZrTe$_5$. The positive slope of $R_{xy}(H)$ as T ≥ 200 K demonstrates that hole carries band dominates. Even though the resistance peak is completely suppressed, the sign reversal still occurs just near the initial $T_P$ (~135 K) for the intrinsic ZrTe$_5$. We believe the similar mechanism of temperature induced Lifshitz transition [26] is still valid to explain the phenomena.

To quantitatively understand the transport properties, Hall conductivity, $\sigma_{xy} = \rho_{xy}/(\rho_{xx}^2 + \rho_{xy}^2)$, was extracted and fitted with the two-carrier model [42],

$$\sigma_{xy} = eB\left(\pm\frac{n_1\mu_1^2}{1+\mu_1^2B^2} \pm \frac{n_2\mu_2^2}{1+\mu_2^2B^2}\right),$$

where $n_{1,2}$ and $\mu_{1,2}$ are the density and mobility of carriers, the plus (minus) sign represents the hole (electron) carriers. Figure 3(a) and (b) show the results obtained for K$_{0.55}$ZrTe$_5$. There are two kinds of electron carriers at $T <$ ~160K. The electron carrier densities n for electron carriers 1 (e1) and 2(e2) is 1.40×10$^{17}$ m$^{-3}$ and 2.37×10$^{19}$ m$^{-3}$ at $T =$ ~20 K, and the mobility u for e1 and e2 are 2.28 m$^2$V$^{-1}$s$^{-1}$ and 0.11 m$^2$V$^{-1}$s$^{-1}$ at T = ~20 K. In comparison with intrinsic ZrTe$_5$, the carrier density is about five orders of magnitude lower, but the mobility has no significant change, consistent with the semiconducting behavior observed in our *R-T* measurement. At $T \approx$ ~160 K, the hole carrier starts to appear. The mobility of hole carriers is greater than that of electron carriers, but the density of hole carriers is smaller, thus corresponding to an initial positive slope in $\rho_{xy}$ (*H*) followed by a change to negative, consistent with previous studies [44]. At higher temperature, two hole carriers dominate, and meanwhile the electron carriers disappear. It is noteworthy that the temperature dependence of the mobility of hole carriers is similar to that observed in Flux-grown ZrTe$_5$ [44]. The observation of carrier type reversal as temperature increases support the Lifshitz transition. Besides the electron-hole pocket competition, the carrier mobility and density may also play a role in the formation of the resistance peak, which has been proposed in the previous study for ZrTe$_5$ nanosheets [42].

As reported previously [44], the flux ZrTe$_5$ single crystal showed the semiconducting-like behavior as well, which was ascribed to the decrease of Te vacancies. In our K intercalated $ZrTe_5$, a difference mechanism should apply, because 1) the location of resistance peak is not changed but the intensity is greatly suppressed, and 2) the electron – hole carrier type transition is still observed between 120 K – 160 K. If only taking into account the electron doping effect, the Fermi level should move upward, and thus an opposite phenomenon should be observed, i.e., the sample should be more metallic behavior at low temperature and the resistance peak should move to higher temperature. Therefore, it suggests that the electron doping effect may play a negligible role.

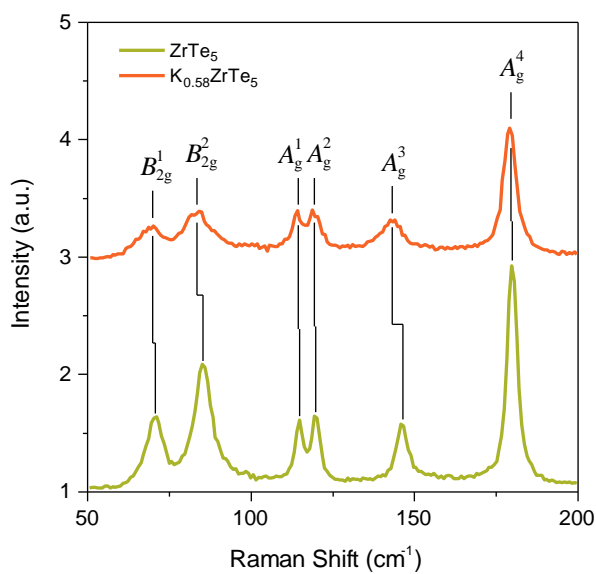

**Figure 4** (Color online) Raman spectra taken on the $ZrTe_5$ single crystal before and after K (x = ~0.58) intercalation. For clarity of viewing the shift, the peaks are marked by vertical lines.

To further characterize the structural change upon K intercalation, Raman spectroscopy measurement is performed for $ZrTe_5$ before and after intercalation, as shown in Fig. 4. According to previous reports [50-52], two $B_{2g}$ modes and four $A_g$ modes can be excited in the range between 50 cm$^{-1}$ and 200 cm$^{-1}$ when the laser vertically irradiates the *ac*-plane. Accordingly, the characteristic peaks at 71.0 cm$^{-1}$ and 85.1 cm$^{-1}$ can be assigned to the $B_{2g}$ modes, and the other four peaks at 114.8 cm$^{-1}$, 119.4cm$^{-1}$, 145.8cm$^{-1}$ and 179.8 cm$^{-1}$ to the $A_g$ modes, as labeled in Fig. 4. Upon K intercalation, the $A_g^3$ mode at 145.8 cm$^{-1}$ makes a clear red shift to 143.3cm$^{-1}$, while the $A_g^4$ mode at 179.8 cm$^{-1}$ is almost unchanged. Such a spectra shift clearly suggests that the lattice constants are changed upon intercalation. According to the previous high pressure study [40], the application of pressure makes the Raman peak at 145.8cm$^{-1}$ blue-shifted, and simultaneously results in the compressed interlayer space. It is thus reasonable to expect in the opposite way that the red-shift of Raman peak, as observed in K$_{0.58}$ZrTe$_5$, is induced by the enlarged interlayer space and weakened interlayer coupling. Therefore, the atomic intercalation plays a "tensile pressure"-like role to enlarge the lattice constant along b axis. Similar intercalation induced lattice expansion has also been reported in some other layered materials [53, 54]. However, in the recent study of K intercalated WTe$_2$ [47], there is no detectable lattice expansion. The reason is not clear why the intercalation induces lattice expansion in some materials, but not for others.

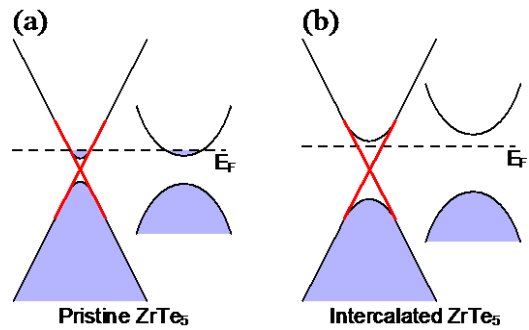

**Figure 5** (Color online) Schematic illustration of the band structures for pristine ZrTe$_5$ (a) and intercalated K$_x$ZrTe$_5$ (b). The red lines represent the topological edge states. For pristine ZrTe$_5$, Fermi level crosses the conduction band at low temperature due to the natural n type doping. After intercalation, the band gap is enlarged and thus the Fermi level is located within the gap region.

Combined with all the above-mentioned results and analysis, a reasonable mechanism can be drawn to explain the observed phenomena, as illustrated in Fig. 5. The pristine ZrTe$_5$ single crystal grown by CVT method is an electron-doped semiconductor which exhibits metallic behaviors at low temperatures, the abnormal resistance peak is caused by temperature-induced Lifshitz transition, as illustrated in Fig. 5(a). Intercalation of K into ZrTe$_5$ applies an effective "negative pressure" and induces the lattice expansion especially in the *b*-axis. Assuming the pristine ZrTe$_5$ is Dirac semimetal or weak TI, such a lattice expansion will definitely enlarge the band gap. As a result, competition between the lattice expansion and electron doping makes the Fermi level located in the band gap region, as shown in Fig. 5(b), and thus an insulating behavior at low temperature.

**Conclusion**

In conclusion, the liquid ammonia method is employed to intercalate alkali metal K into single crystal ZrTe$_5$, inducing the lattice expansion and then enlarging the band gap. As a result, the ZrTe$_5$ is tuned from semimetal-like to insulating state at low temperature. Hall sign reversal indicates the presence of Lifzhitz transition. Positive magnetoresistance effect follows a $R \sim \sqrt{B}$ dependence at low temperature, which can be attributed to the weak antilocalization effect. This work provides an ideal material platform of insulating ZrTe$_5$ for further explorations, particularly regarding to the transport properties of the topological edge states.

**Acknowledgement:**

This work was supported by the Ministry of Science and Technology of China (Grants Nos. 2014CB921103, 2015CB921203), the National Natural Science Foundation of China (Grants Nos. 11774149, 11790311, 11674157, 11674154, 51032003, 1171101156, 11374149, and 11374140).